\documentclass[prl,superscriptaddress,nofootinbib,floats,twocolumn,longbibliography]{revtex4-2}

\usepackage{amssymb,amsmath}
\DeclareMathAlphabet\mathbfcal{OMS}{cmsy}{b}{n}
\usepackage{cmap}
\usepackage{graphicx} 
\usepackage{bm}
\usepackage{color}
\usepackage{blindtext}
\usepackage{hyperref}
\usepackage{listings}
\usepackage{booktabs}
\usepackage{multirow}
\usepackage{amsmath}
\usepackage{mathrsfs}
\usepackage{braket}
\usepackage{mathtools}
\usepackage{sidecap}
\usepackage{dsfont}
\usepackage{tabularx}
\usepackage{mhchem}

\begin{document}

\title{Twisted multilayer moir\'e water waves topologically robust to disorder}

\author{Zhiyuan Che$^\dagger$}
\affiliation{State Key Laboratory of Surface Physics, Key Laboratory of Micro- and Nano-Photonic Structures (Ministry of Education), and Department of Physics, Fudan University, Yangpu District, Shanghai, 200433, China}


\author{Julian Schwab$^\dagger$}
\affiliation{4th Physics Institute, Research Center SCoPE, and Integrated Quantum Science and Technology Center, University of Stuttgart, Germany}
\affiliation{Centre for Disruptive Photonic Technologies, School of Physical and Mathematical Sciences \& The Photonics Institute, Nanyang Technological University, Singapore 637371, Singapore}

\author{Yi Zhang}
\author{Junyi Ye}
\author{Cheng Cheng}

\affiliation{State Key Laboratory of Surface Physics, Key Laboratory of Micro- and Nano-Photonic Structures (Ministry of Education), and Department of Physics, Fudan University, Yangpu District, Shanghai, 200433, China}

\author{Lei~Shi$^*$}

\affiliation{State Key Laboratory of Surface Physics, Key Laboratory of Micro- and Nano-Photonic Structures (Ministry of Education), and Department of Physics, Fudan University, Yangpu District, Shanghai, 200433, China}

\affiliation{Institute for Nanoelectronic devices and Quantum computing, Fudan University, Shanghai 200438, China}
\affiliation{Collaborative Innovation Center of Advanced Microstructures, Nanjing University, Nanjing 210093, China}
\affiliation{Shanghai Research Center for Quantum Sciences, Shanghai 201315, China}

\author{Yijie Shen$^*$}
\affiliation{Centre for Disruptive Photonic Technologies, School of Physical and Mathematical Sciences \& The Photonics Institute, Nanyang Technological University, Singapore 637371, Singapore}

\affiliation{School of Electrical and Electronic Engineering, Nanyang Technological University, Singapore 639798, Singapore}

\author{Harald Giessen$^*$}
\affiliation{4th Physics Institute, Research Center SCoPE, and Integrated Quantum Science and Technology Center, University of Stuttgart, Germany}

\author{Jian Zi$^*$}
\affiliation{State Key Laboratory of Surface Physics, Key Laboratory of Micro- and Nano-Photonic Structures (Ministry of Education), and Department of Physics, Fudan University, Yangpu District, Shanghai, 200433, China}
\affiliation{Institute for Nanoelectronic devices and Quantum computing, Fudan University, Shanghai 200438, China}
\affiliation{Collaborative Innovation Center of Advanced Microstructures, Nanjing University, Nanjing 210093, China}
\affiliation{Shanghai Research Center for Quantum Sciences, Shanghai 201315, China}

\begin{abstract}
Moir\'e patterns, stacking and twisting multilayer periodic lattices into superlattices, have become cornerstones of many physical systems from condensed matter to wave phenomena, but have never been properly studied in water waves. Here, we demonstrate twisted multilayer moir\'e water surface waves carrying robust skyrmionic topologies. Using a custom water tank of circular multi-channel phased array, we precisely generate water-wave skyrmion lattices and superimpose them into moir\'e superlattices with higher-order topological textures, e.g., skyrmion bags and clusters, programmed via the twist angle. We also quantitatively compare the topological robustness of bilayer and trilayer configurations under spatiotemporal perturbations. The trilayer moiré superlattices exhibit more enhanced stability, stronger field localization and energy concentration than the bilayer. Our work establishes water waves as a macroscopic, tunable, and visually accessible platform for moir\'e physics, towards robust particle manipulation and classical analogues of topological quantum phenomena.
\end{abstract}

{
\let\clearpage\relax
\maketitle
}

\def\thefootnote{$\dagger$}\footnotetext{These authors contributed equally to this work}

\def\thefootnote{*}\footnotetext{Email: lshi@fudan.edu.cn, yijie.shen@ntu.edu.sg, giessen@pi4.uni-stuttgart.de, jzi@fudan.edu.cn}

\vspace{-0.5cm}


\textbf{\emph{Introduction}}--Topology, the study of properties preserved under continuous deformation, provides a powerful framework for physics \cite{nakahara2018}. 
One key finding from topology is the classification of field configurations. In this context, topological wave forms \cite{nye1974,nye1987,bauer2015}, such as skyrmions \cite{skyrme1961,Pfleiderer2010,Foster2019,leslie2009,cruz2007,blaauwgeers2000,Ge2021, Muelas-Hurtado2022, Yang2015,Shen2023reviewnatphot, Tsesses2018, Davis2020,RN3166, RN3164, Dev2025}, have become of particular interest. 
An intriguing avenue for engineering topological complexity is through moiré patterns, which are formed by superimposing periodic structures with a relative twist \cite{cao2018, andrei2021}. Recently, moiré lattices enabled superlattices harboring skyrmion bags: multiple skyrmions confined within a larger one of opposite polarity \cite{Schwab2025, Foster2019}.

Topological textures exhibit remarkable robustness, preserving global topology despite local distortions. In free-space optics, skyrmionic beams maintain their topological structure even when subjected to severe atmospheric turbulence \cite{guo2025topological} or propagation through complex scattering media \cite{Wang2024complexmedia}. This stability also extends to dynamical \cite{Wu2025} and quantum skyrmion systems \cite{guo2025topological,Ornelas2025noise, Wang2025storage} in the far and near-field regime \cite{Deng2022}.
In addition, high-order states, such as skyrmion bags, also exhibit robustness against deviations in the superlattice twist angle \cite{Schwab2025}, structural disorder \cite{Schwab2025nanoph}, or perturbations and obstacles during free-space propagation \cite{Dev2025,guo2025self}.

Lately, it has been demonstrated that the inherently vectorial displacement field of water waves closely resembles optical surface waves \cite{Smirnova2024, Wang2025}, such as surface plasmon polaritons. Hence, the complex motion of interfering water waves also enables the realization of complex topological field structures, such as skyrmions \cite{Wang2025}.

In this work, we introduce and experimentally demonstrate water-wave skyrmion lattices and their bilayer and trilayer moiré superlattices. 
Compared to their counterparts in evanescent surface waves, our superlattices feature visual accessibility with a large field of view and fast, accurate tunability.
This allows us to apply an experimental method \cite{Schwab2025nanoph} to quantify the topological robustness of vector textures against random wave disturbances and study their energy localization. 
Our results expand the toolbox of water waves and their ability to manipulate the dynamics of floating objects. The work also highlights water waves as a versatile platform for investigating topological properties of structured wave systems and opens new directions for robust classical analogues of topological quantum materials \cite{Xu2024}.

\begin{figure}[t]
	\centering
	\includegraphics[width=1\linewidth]{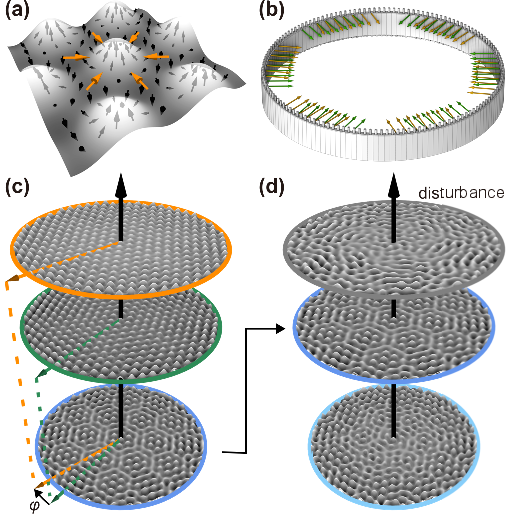}
	\caption{\textbf{Moir\'e skyrmion superlattices in water waves.} 
		\textbf{(a)} Conceptual illustration of a skyrmion lattice formed in the three-dimensional displacement field  \(\mathbfcal{R}(x,y,t)\) via interference of six plane waves in a hexagonal arrangement (orange arrows). 
		\textbf{(b)} Experimental scheme. Multiple acoustic sources in a ring-shaped cavity excite water waves, forming two twisted hexagonal skyrmion lattices (orange, green arrows).
        Their superposition creates the moir\'e superlattice. 
		\textbf{(c)} Measured vertical displacement field of two skyrmion lattices (orange, green) with a relative twist angle $\varphi=13.17^\circ$. Their moiré superlattice (blue) is obtained by simultaneously exciting both skyrmion lattices. Following this principle, trilayer superlattices can be constructed by adding a third twisted skyrmion lattice. 
		\textbf{(d)} The topological robustness is probed by introducing a controlled perturbation to the moiré skyrmion superlattice.
	}
	\label{fig1}
\end{figure}


\textbf{\emph{Water-wave moiré superlattices}}--To generate skyrmion lattices [Fig.~\ref{fig1}(a)] and moiré superlattices, we employ a circular array of 192 micro-apertures [Fig.~\ref{fig1}(b)] divided into six symmetric sectors. Each sector comprises 32 independently controlled speakers. Precise modulation of their phase and amplitude produces a hexagonal standing-wave pattern (Supplementary Fig.~S1), where each unit cell contains a skyrmion. 
Superimposing two or three such lattices with relative twist angles
creates bilayer or trilayer moiré superlattices [Fig.~1(c)].
Experiments were conducted in an $80 \times 80\,{\rm cm}^2$ tank (depth $h=2.5\,$cm) at $f_0 = 8.02\,$Hz ($\lambda = 3$~cm), satisfying the deep-water condition $\tanh(kh) \simeq 1$.

\begin{figure*}[t]
	\centering
	\includegraphics[width=0.85\linewidth]{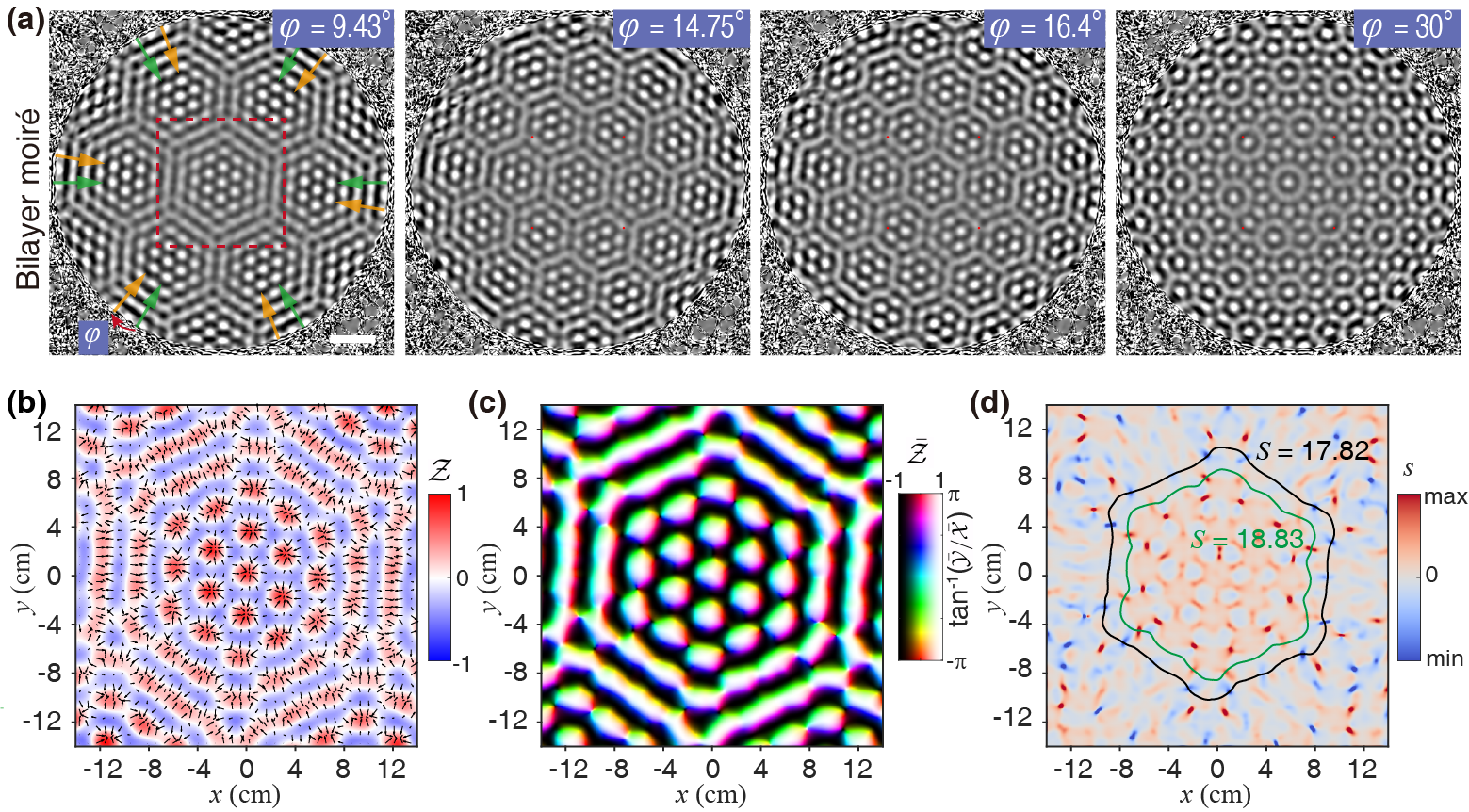}
	\caption{
		\textbf{Bilayer moiré skyrmion superlattices in water waves.}
		\textbf{(a)} Measured displacement field $\mathcal{Z}(x,y,t=0)$ for different interlayer twist 
        angles $\varphi$. 
		\textbf{(b-d)} Magnified view of the red-boxed region in (a). \textbf{(b)} Vertical field $\mathcal{Z}$ with reconstructed in-plane displacement vectors $(\mathcal{X},\mathcal{Y})$ indicated by arrows.
		\textbf{(c)} 3D vector field distribution $\mathbfcal{R}(x,y,t=0)$ encoded by color (in-plane orientation) and brightness (out-of-plane amplitude).
		\textbf{(d)} Skyrmion density distribution. Integration using the boundaries of the skyrmion cluster (green) and the total skyrmion bag (black) yields topological charges of $S_\text{cluster} \simeq 18.83$ and $S_\text{bag} \simeq 17.82$.
	}
	\label{Fig_2}
\end{figure*}

\par Fig.~\ref{Fig_2}(a) presents the measured vertical displacement field \(\mathcal{Z}(x,y,t=0)\) for twisted bilayer moiré superlattices with various twist angles. 
While twist angles of $\varphi = 9.43^\circ$ and $\varphi = 16.4^\circ$ result in periodic superlattices, $\varphi = 14.75^\circ$ and $\varphi = 30^\circ$ yield a quasicrystal superlattice \cite{Schwab2025}. 
The vertical displacement fields are obtained via schlieren imaging and fast checkerboard demodulation \cite{Wildeman2018, SM}. We reconstruct the full vector field \(\mathbfcal{R}(x,y,t)\) by deriving its horizontal components from the in-plane gradient \((\mathcal{X},\mathcal{Y}) = k^{-1} \bm{\nabla}_2 \mathcal{Z}\) \cite{Smirnova2024, Wang2025, Shi2019NSR, Muelas2022PRL}. The full vector fields reveal textures, identified as skyrmion bags and clusters \cite{Foster2019, Schwab2025}, that go beyond previously reported water-wave skyrmions with a unit topological charge of \(S=1\) \cite{Wang2025}. At 
\(\varphi = 9.43^\circ\) [Fig.~\ref{Fig_2}(b-d)], we observe a skyrmion bag that encloses a cluster of 19 individual skyrmions with total skyrmion number \(S_\text{cluster}=19\). This cluster is surrounded by a larger skyrmion of opposite polarity (\(S=-1\)). As a result, the total skyrmion bag yields a topological charge of \(S_{\text{bag}} = 18\).
The local topology of the measured vector field is quantitatively analyzed using the skyrmion number density
$s = \bar{\mathbfcal R} \cdot (\partial_x \bar{\mathbfcal R} \times \partial_y \bar{\mathbfcal R})$, where $\bar{\mathbfcal R} = \mathbfcal{R}(x,y,t)/|\mathbfcal{R}(x,y,t)|$ [Fig.~\ref{Fig_2}(d)]. The topological charge (skyrmion number) is given by \(S = (1/4\pi) \iint s \, dx \, dy\). The integration is performed over the region enclosed by the minimal (maximal) contour of \(\mathcal{Z}\) for the skyrmion cluster (bag), indicated by the green (black) curve in Fig.~\ref{Fig_2}(d). By integrating, we find \(S_\text{cluster} \simeq 18.83\) and \(S_\text{bag} \simeq 17.82\), yielding near-integer values in agreement with theoretical expectations.
By manipulating the twist angle and the center of rotation, the topological charge of the skyrmion bags can be controlled. For instance, at \(\varphi = 16.4^\circ\) we observe a skyrmion bag with $S_\text{bag}=6$ comprising $S_\text{cluster}=7$ skyrmions (Fig. S3). 

\begin{figure*}[t]
	\centering
	\includegraphics[width=0.85\linewidth]{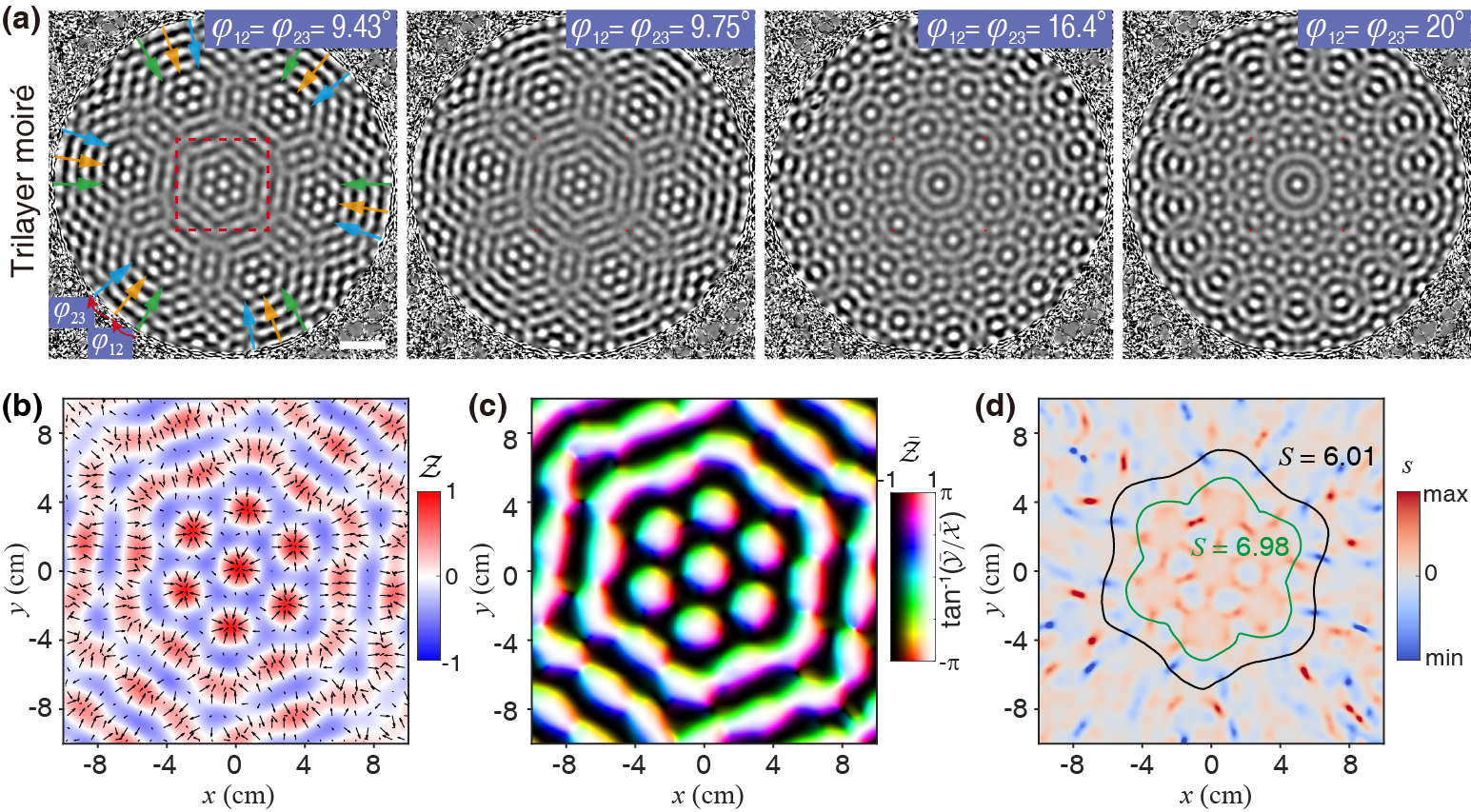}
	\caption{
		\textbf{Trilayer moiré skyrmion superlattices in water waves.}
		\textbf{(a)} Measured vertical displacement field $\mathcal{Z}(x,y,t=0)$ for trilayer moiré superlattices with twist angles $\varphi_{12} = \varphi_{23}. $
		\textbf{(b-d)} Magnified view of the red-boxed region in (a). \textbf{(b)} Vertical field $\mathcal{Z}$ with reconstructed in-plane displacement fields $(\mathcal{X},\mathcal{Y})$ indicated by arrows.
		\textbf{(c)} 3D vector field distribution $\mathbfcal{R}(x,y,t=0)$ encoded by color (in-plane orientation) and brightness (out-of-plane amplitude).
		\textbf{(d)} Skyrmion density distribution. 
        Contours denote integration boundaries yielding topological charges $S_\text{cluster} \simeq 6.98$ and $S_\text{bag} \simeq 6.01$.
	}
	\label{Fig_3}
\end{figure*}

Building on the bilayer configuration, the flexibility of our experimental platform allows for the addition of a third hexagonal lattice, thereby creating trilayer moiré superlattices, with two distinct twist angles between the first and second lattice ($\varphi_{12}$), as well as between the second and third lattice ($\varphi_{23}$) \cite{Park2021,Schwab2025nanoph}. 
Fig. \ref{Fig_3}(a) presents the measured vertical displacement field \(\mathcal{Z}(x,y,t=0)\) of trilayer moiré superlattices. 
The trilayer configuration yields much larger super unit cells, harboring multiple skyrmion bags and clusters, so that the periodicities at $\varphi_{12} = \varphi_{23} = 9.43^\circ$ can not be observed.
For correctly chosen twist angles, the bilayer and trilayer superlattices can harbour the same skyrmion bag configurations \cite{Schwab2025nanoph}. We experimentally demonstrate this for a skyrmion bag with $S_\text{bag}=6$ harboring $S_\text{cluster}=7$ skyrmions at twist angles \(\varphi_{12} = \varphi_{23} = 9.43^\circ\) depicted in Fig.~3(b-d). 
Integration of the skyrmion density gives topological charges of \(S_\text{cluster} \simeq 6.98\) for the cluster and \(S_\text{bag} \simeq 6.01\) for the total skyrmion bag. 
In the Supplementary Materials \cite{SM}, we present the results for unequal successive twist angles \(\varphi_{12} \neq \varphi_{23}\) and further demonstrate that we achieve programmable generation of skyrmion bags and clusters with targeted topological charges by varying the interlayer twist angles and rotation centers.
This extends the controlled creation of such textures, recently demonstrated for optical skyrmions in bilayer systems \cite{Schwab2025}, to a more versatile macroscopic platform. 


\begin{figure}[t]
	\centering
	\includegraphics[width=1\linewidth]{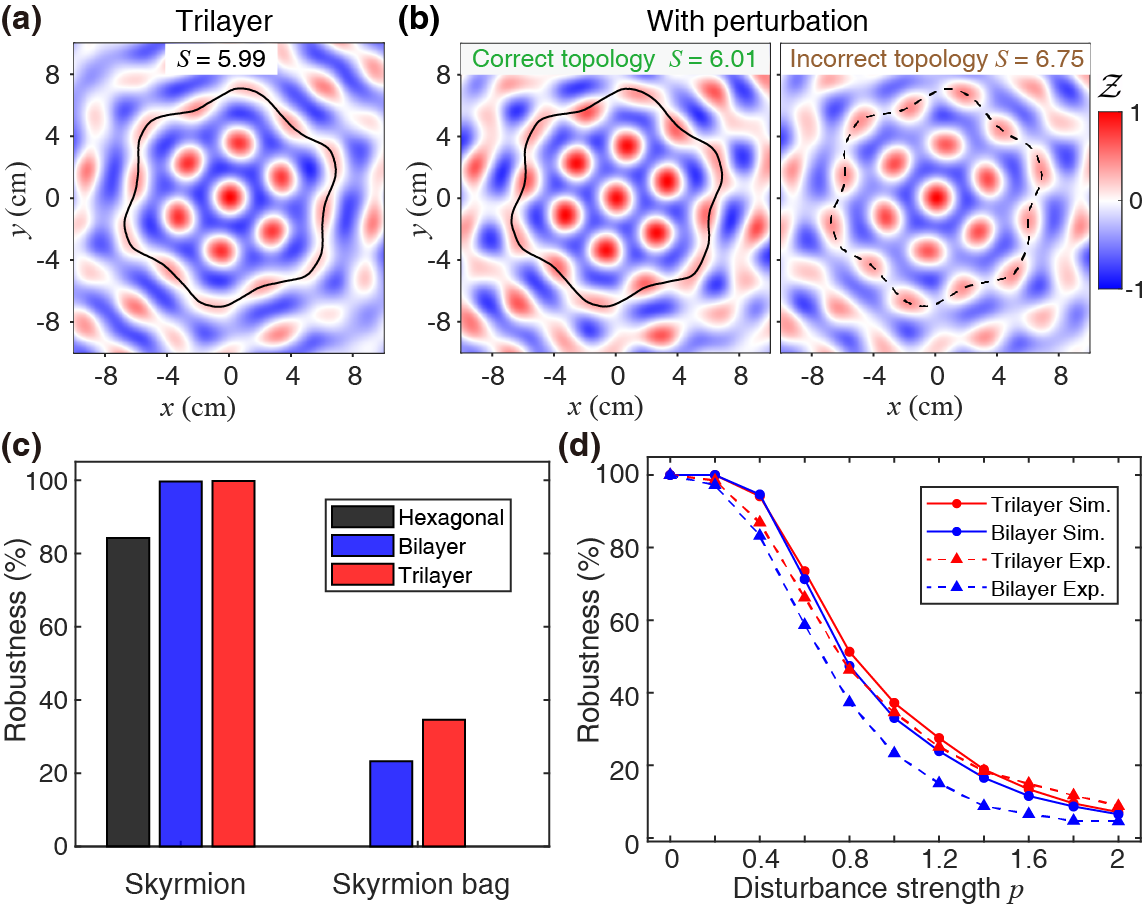}
	\caption{
		\textbf{Enhanced topological robustness in moiré skyrmion superlattices under perturbation.}
		\textbf{(a)} A skyrmion bag in an unperturbed trilayer superlattice (\(S_\text{bag} \simeq 5.99\)). \textbf{(b)} Response to perturbation (\(p=1\)): a stable texture with topological charge (\(S_\text{bag} \simeq 6.01\), left) and a distorted case yielding a non-integer charge (\(S_\text{bag} \simeq 6.75\), right).
		\textbf{(c)} Layer-dependent robustness. Statistical robustness of unit-charge skyrmions (\(S=1\)) under fixed perturbation (\(p=1\)). The monolayer lattice (black) exhibits minimal robustness. Bilayer (blue, \(\varphi_{12}=14.75^\circ\)) and trilayer (red, \(\varphi_{12}=\varphi_{23}=9.75^\circ\)) configurations achieve near-perfect robustness, with trilayers consistently superior.
		\textbf{(d)} Robustness versus perturbation strength. Measured (dashed) and simulated (solid) robustness \(\mathscr{R}(p)\). Trilayer superlattices (red) exhibit superior stability across all \(p\).
	}
	\label{fig:robustness}
\end{figure}

\begin{figure}[t]
	\centering
	\includegraphics[width=1\linewidth]{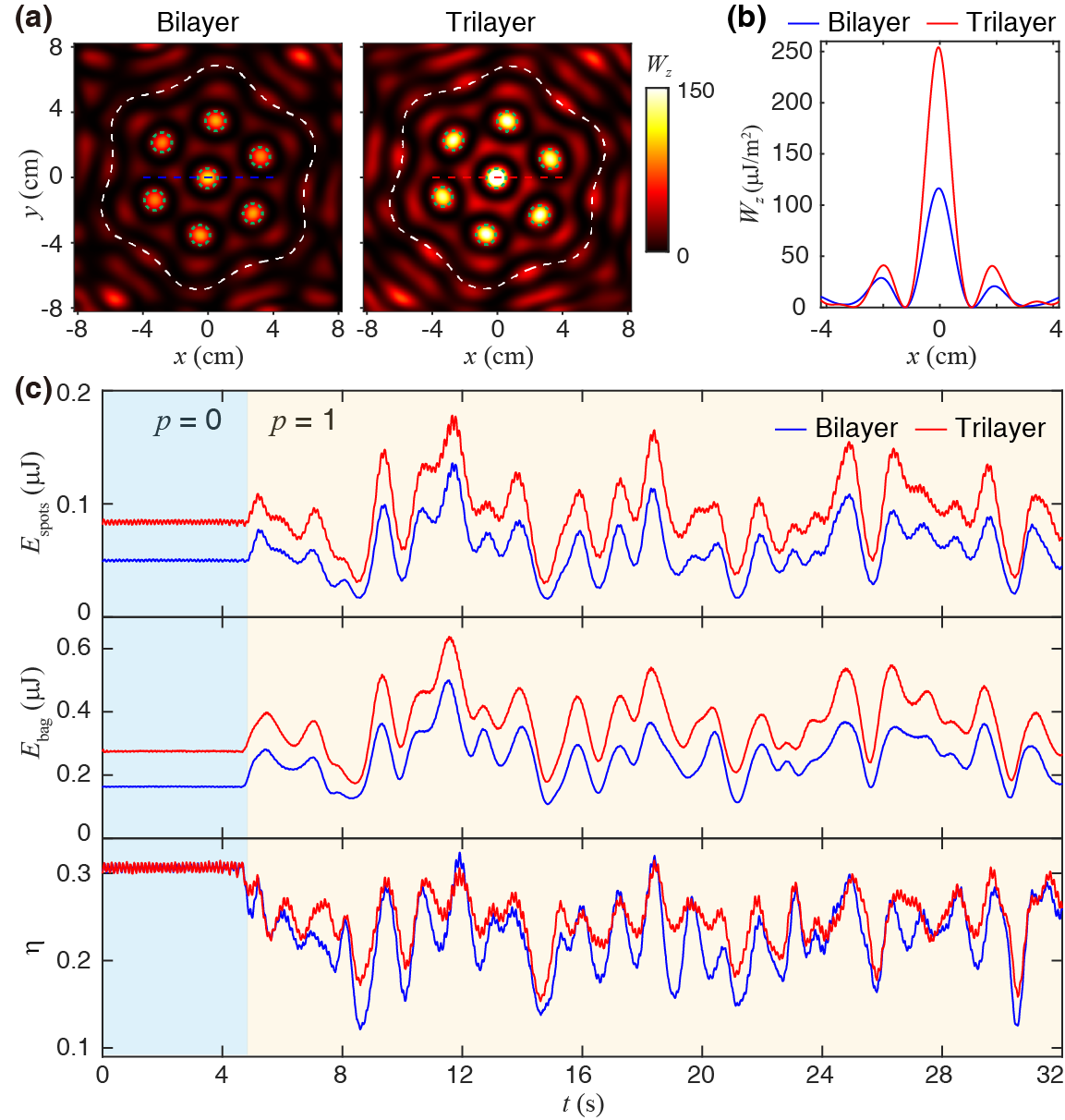}
	\caption{
		\textbf{Enhanced energy localization in moiré skyrmion superlattices.}
		\textbf{(a)} Spatial distribution of vertical energy density \(W_z\) reveals localized hotspots. \textbf{(b)} Cross-sectional profiles show the trilayer’s peak intensity exceeds the bilayer’s by more than a factor of two.
		\textbf{(c)} Dynamics of $E_{\text{spots}}$, $E_{\text{bag}}$, and \(\eta(t)\) under sustained perturbation (\(p=1\)). The trilayer (red) maintains a higher \(\eta(t)\) than the bilayer (blue), demonstrating stronger energy localization in the hotspots.
	}
	\label{fig:localization}
\end{figure}

\textbf{\emph{Topological robustness of skyrmion bags}}--For practical applications, the utility of complex topological textures relies on their resilience to external perturbations and their ability to confine energy. Theoretical studies suggest that increasing the number of moiré layers can enhance topological protection \cite{Schwab2025nanoph}. To test this hypothesis experimentally, we perform a quantitative comparison of robustness and energy localization between bilayer and trilayer moiré skyrmion superlattices, focusing on (1) the stability of topological numbers under controlled disorder, and (2) the degree of energy localization within the non-trivial texture.

To quantify robustness, a controlled perturbation is introduced by superimposing pseudo-random signals onto the 32 independent driving channels of the phased array [Fig.~1(c)]. The amplitude, phase, and frequency for each channel follow a Gaussian distribution, generating a spatiotemporal perturbation field, \(\mathcal{Z}_d(x,y, t)\). The total field \(\mathcal{Z}(x,y, t)\), for a perturbed bilayer or trilayer superlattice, is then given by the superposition of the unperturbed field \(\mathcal{Z}_0(x,y, t)\) and the scaled perturbation:
\begin{equation}
	\mathcal{Z}(x,y, t) = \mathcal{Z}_0(x,y, t) + p \frac{|\mathcal{Z}_0(x,y, t)|}{|\mathcal{Z}_d(x,y, t)|}\mathcal{Z}_d(x,y, t),
	\label{eq:perturbance}
\end{equation}
where \(p\) is a dimensionless global perturbation strength. In this perturbed field configuration

Topological robustness, \(\mathscr{R}\), is quantified as the system's ability to retain its ideal topological charge \(S_{\text{ideal}}\) under perturbation. A configuration is deemed topologically correct if its measured topological charge \(S\) satisfies:
\begin{equation}
	1 - \varepsilon < \frac{S}{S_{\text{ideal}}} < 1 + \varepsilon, \label{eq:Robustness_tolerance}
\end{equation}
where \(\varepsilon\) is a small tolerance [Fig.~\ref{fig:robustness}(b)]. The robustness metric is then statistically defined as:
\begin{equation}
	\mathscr{R} = \frac{N_{\text{correct}}}{N_{\text{total}}},
	\label{eq:Robustness}
\end{equation}
where \(N_{\text{correct}}\) is the number of measurement frames with correct topology and \(N_{\text{total}}\) is the total number of frames analyzed.

To quantitatively compare the topological robustness, we generated bilayer and trilayer moiré superlattices, each engineered to host identical, well-defined topological charges (\(S_{\text{ideal}}\)), and operated at their respective twist angles for optimal stability. For unit-charge skyrmions (\(S_{\text{ideal}} = 1\)), realized at \(\varphi=30^\circ\) [bilayer; \text{Fig.}~\ref{Fig_2}(a), right] and \(\varphi_{12}=\varphi_{23}=20^\circ\) [trilayer; \text{Fig.}~\ref{Fig_3}(a), right], both multilayer systems exhibited near-perfect robustness (\(\sim 99.6\text{--}99.8\%\)) under a perturbation strength of \(p=1\), significantly exceeding the 84.2\% robustness of a monolayer hexagonal lattice [\text{Fig.}~\ref{fig:robustness}(c), left]. Although the trilayer's advantage here is marginal, a decisive enhancement emerges for higher-order textures. For a skyrmion bag with \(S_{\text{cluster,ideal}} = 7\) and \(S_{\text{bag,ideal}} = 6\), both of its topological numbers need to fulfill eq. \eqref{eq:Robustness_tolerance}. The skyrmion bags are realized at \(\varphi=14.75^\circ\) in the bilayer and \(\varphi_{12}=\varphi_{23}=9.75^\circ\) in the trilayer configuration. Notably, the trilayer's robustness (34.6\%) substantially surpasses that of the bilayer (23.3\%) at the same \(p=1\) [Fig.~\ref{fig:robustness}(c), right], underscoring its superior capability to stabilize complex topological states.

This enhanced stability in trilayer systems is systematic. As shown in \text{Fig.}~\ref{fig:robustness}(d), the measured robustness \(\mathscr{R}(p)\) for trilayer superlattices (red dashed curve) consistently exceeds that of bilayer systems (blue dashed curve) across the full range of perturbation strengths \(p\). The experimental trends are in good qualitative agreement with numerical simulations (solid curves), with the systematically lower experimental values attributable to unavoidable environmental disturbances in the physical implementation. Collectively, these results demonstrate that trilayer moiré engineering provides a distinctly more robust platform for sustaining high-charge topological textures against perturbations.

A primary reason for the superior topological protection of the trilayer moiré superlattices is the significantly enhanced energy localization compared to the bilayer systems.
This capability arises from an expanded momentum-space structure: trilayer configurations support interference among 18 distinct wavevectors, surpassing the 12 waves of the bilayer system. The resulting, more complex interference pattern enables the formation of sharper and more intense energy hotspots.

The enhanced localization directly translates to steeper energy gradients, a critical factor for amplifying wave-matter interactions. In the water-wave system, the kinetic energy surface density (vertical) is given by \cite{Peskin2010,Bliokh2022SA}:
\begin{equation}
	W_z(x,y) = \frac{\rho \omega^2}{4k}  |Z(x,y)|^2,
\end{equation}
where $\rho$ is the fluid density, $\omega$ is the angular frequency, and $Z$ is the complex vertical displacement field \cite{SM}.

Under the same driving conditions, the spatial distribution of $W_z$ reveals a pronounced contrast [Fig.~\ref{fig:localization}(a)]. Energy in the trilayer superlattice is concentrated far more effectively, significantly intensifying all seven primary hotspots (marked by green dashed circles), with the most dramatic enhancement occurring at the center. Quantitative cross-sectional profiles [Fig.~\ref{fig:localization}(b)] confirm that the peak energy density in the trilayer structure is more than double that of its bilayer counterpart, validating its superior sub-wavelength focusing capability.

\par This enhanced localization persists dynamically under external perturbation. Fig.~\ref{fig:localization}(c) shows the time evolution of the total energy concentrated in the dominant hotspots (\(E_{\text{spots}}\), top panel) and the total energy confined within the topological texture boundary (\(E_{\text{bag}}\), middle panel) under sustained perturbation strength (\(p=1\)). Both quantities remain consistently higher for the trilayer than for the bilayer throughout the dynamics. To quantify the perturbation-induced energy redistribution, we define the localization ratio \(\eta(t)=E_{\text{spots}}/E_{\text{bag}}\). Notably, the trilayer maintains a higher \(\eta(t)\) over time [Fig.~\ref{fig:localization}(c), bottom panel], demonstrating its more robust confinement of energy into hotspots despite disturbances. This persistent localization could be harnessed in the future for robust particle trapping and manipulation \cite{Wang2025}.

\textbf{\emph{Conclusions}}--
Twisted multilayer moiré water waves establish a macroscopic platform for exploring topological physics. By superimposing skyrmion lattices, we generate skyrmion bags—higher-order textures previously restricted to microscopic systems. Our platform makes these vector textures visually accessible and tunable via twist angles and rotation centers. This flexibility allows us to demonstrate that trilayer configurations exhibit enhanced topological robustness against perturbations.

The establishment of water waves as a robust platform for moiré topological physics opens several promising research avenues. The demonstrated energy localization and topological stability can be directly applied for trapping and manipulation of floating sub-wavelength particles with increased stability. Future studies could also explore more complex stacking configurations beyond trilayers to further amplify energy localization and investigate the limits of topological protection against environmental disorders. 
The trapping of multiple floating particles within stable topological lattices on the water-wave surface could also enable the observation of collective particle dynamics. These macroscopic interactions might serve as classical analogues for understanding complex behavior in topological quantum materials \cite{Xu2024}. By scaling these quantum-level phenomena to the visually accessible water surface, this opens a path toward simulating and engineering robust many-body physics in a macroscopic wave system.

\vspace{0.5cm}

\begin{acknowledgements}

{\bf Acknowledgements:}
The authors acknowledge the support of National Natural Science Foundation of China (No. 12234007, No. 12321161645, No. 12221004, Grants No. T2394480 and No. T2394481); National Key Research and Development Program of China (2023YFA1406900, 2022YFA1404800); Science and Technology Commission of Shanghai Municipality (2019SHZDZX01, 23DZ2260100);
Singapore Ministry of Education (MOE) AcRF Tier 1 grants (RG157/23 \& RT11/23), Singapore Agency for Science, Technology and Research (A*STAR) MTC Individual Research Grants (M24N7c0080), and Nanyang Assistant Professorship Start Up grant;
ERC (Complexplas, 3DPrintedoptics); DFG (SPP1391 Ultrafast Nanooptics, CRC 1242 “Non-Equilibrium Dynamics of Condensed Matter in the Time Domain” project no. 278162697-SFB 1242); BMBF (Printoptics); BW Stiftung (Spitzenforschung, Opterial); Carl-Zeiss Stiftung. J.S. acknowledges support from the Vector Stiftung (funding line “MINT-Innovationen”) and Alexander von Humboldt Foundation.

{\bf Author contributions:}
All the authors discussed, interpreted the results, and conceived the theoretical framework. 
Z. C., J. S., Y. S. and L. S. conceived the basic idea of the work. 
Z. C. extended the model to liquid surface waves and designed the numerical calculations. 
Z. C. designed and performed the experiments.
Z. C. and J. S. analyzed the experimental data. Z. C., J. S., and Y. S. wrote the manuscript draft. 
L. S., Y. S., H. G. and J. Z. supervised the research and the development of the manuscript. 
All authors took part in the discussion, revision, and approved the final copy of the manuscript.

Z. C. and J. S. contributed equally to this work.

{\bf Competing interests:} 
The authors declare no competing interests.

{\bf Data and materials availability:} 
All data needed to evaluate the conclusions in the paper are present in the paper and the Supplementary Materials.  
\end{acknowledgements}

\bibliography{library}

\end{document}